\documentclass[seceq,preprint]{ptptex}

\usepackage{graphics}
\preprintnumber[3cm]{%<-- [..]: optional width of preprint # column.
DPNU-03-10\\ hep-ph/0305073}
\title{%        %You can use \\ for explicit line-break
Calculation of Magnetic Penguin Amplitudes \\in $B\to\phi K$ Decays using 
PQCD Approach 
}

\author{%       %Use \scshape  for the family name
Satoshi \scshape{Mishima}\footnote{E-mail: mishima@eken.phys.nagoya-u.ac.jp}
 and 
A. I. \scshape{Sanda}\footnote{E-mail: sanda@eken.phys.nagoya-u.ac.jp}
}

\inst{%         %Affiliation, neglected when [addenda] or [errata]
Department of Physics, Nagoya University, Nagoya 464-8602, Japan
}

\recdate{May 8, 2003}%            %Editorial Office will fill in this.

\abst{%         %this abstract is neglected when [addenda] or [errata]
New physics contributions to $B$ decays often arise through
chromo-magnetic penguin operators. To look for new physics effects in $B$
decays, it is useful to be able to estimate the hadronic matrix elements
for the chromo-magnetic operator. We compute this contribution to
$B\to\phi K$ decays using PQCD methods. It is shown that, if the Wilson
coefficient of the new physics is same order of magnitude as that of the
Standard Model, 
this operator gives a non-negligible contribution compared to that of the
Standard Model (about 30\%). We also investigate the value of $q^2$,
which is the momentum transferred by the gluon in the chromo-magnetic
penguin operator. We find that the expectation value $\langle
q^2\rangle$ is approximately $M^2_B/4$, in agreement with a naive
guess. 
This result, however, is very sensitive to the scale dependence of the
Wilson coefficient. We also show that the matrix element for the
chromo-magnetic penguin operator is independent of the choice of energy
scale to a very good approximation. 
}

\begin{document}

\maketitle

%-------------------------------------------------------------------%
%
%                  Introduction
%
%-------------------------------------------------------------------%
\section{Introduction}

$B\to \phi K_S$ decay plays an important role in the search for new
physics beyond the Standard Model (SM),
as this decay is a pure penguin process. In the SM, the time
dependent {\it CP} asymmetry for $B\to \phi K_S$ decay is the same as
that of $B\to J/\psi K_S$ 
decay, which is a tree dominant process. 
Any difference between them would be a signal for physics beyond the SM.
Recently, the BaBar~\cite{Aubert:2002nx} and Belle~\cite{Abe:2002bx}
collaborations reported {\it CP} asymmetry in $B\to \phi K_S$ decay. 
If the present trend continues with additional data, the results suggest
a deviation from the SM prediction.  

In the effective Hamiltonian approach,\cite{Buchalla:1995vs} \ the
Hamiltonian is expressed as the 
convolution of local operators and the Wilson coefficients, which can
include contribution from new physics.
The effects of new physics change the Wilson coefficients.
In many cases, the Wilson coefficient for the chromo-magnetic operator
is most sensitive to new
physics.\cite{Keum:1999vt,Moroi:2000tk,Causse:2002mu, 
Kane:2002sp,Harnik:2002vs,Baek:2003kb,Khalil:2003bi} \ 
The factorization approximation (FA) is
often used in computing the hadron matrix elements. 
However, it is difficult to calculate this contribution in the FA,
because the magnitude of $q^2$ of the virtual gluon in the
chromo-magnetic penguin operator is unknown.
In the FA, $q^2$ is often assumed to be $M_B^2/4$ or $M_B^2/2$.
Then, the result is sensitive to the assumed value of
$q^2$.\cite{Ali:1997nh}

In this paper, we calculate the chromo-magnetic penguin using the 
perturbative QCD (PQCD) approach within the SM.
We can predict not only the factorizable contributions but also
non-factorizable contributions and annihilation contributions, which
cannot be calculated in the FA.
Furthermore, we can calculate strong phases, so that it is possible to
predict {\it CP} asymmetries.
PQCD has been applied to some exclusive decay modes,
i.e. $K\pi$,\cite{Keum:2000wi} $\pi\pi$,\cite{Lu:2000em}  
$KK$,\cite{Chen:2000ih}   
$K\eta'$,\cite{Kou:2001pm}  $D_sK$,\cite{Lu:2002iv}  
$\rho(\omega)\pi$,\cite{Lu:2000hj}   
$\phi K$,\cite{Mishima:2001pp,Chen:2001pr}   
$K^*\pi$\cite{Keum:2002qj} and $\phi K^*$,\cite{Chen:2002pz} and the
results are consistent with experimental data. 
They were calculated to leading order in $\alpha_s$ in the PQCD
approach. Therefore, the chromo-magnetic penguin was not included in the
computation of the branching ratio for the $B\to\phi K$ decay. 
Obviously, there are many other higher-order diagrams that must be
considered simultaneously if we are to add the magnetic penguin
term to the SM computation.
To estimate the new physics contribution, the magnetic penguin
contribution can be singled out, as it is most sensitive to new physics. 
In this study, we also analyze the FA computation of the color magnetic
moment using PQCD as a guide. 

The outline of this paper is as follows. First, we present the effective
Hamiltonian and the chromo-magnetic penguin operator. Then, 
we derive magnetic penguin contributions for hadronic two-body
decays. Next, we calculate the chromo-magnetic penguin contributions in
$B\to\phi K$ decays using the PQCD approach and we give the numerical
result. We also derive the distribution of $q^2$.
Finally, we summarize this study.

%-------------------------------------------------------------------%
%
%                Chromo-magnetic Penguin Operator
%
%-------------------------------------------------------------------%
\section{Chromo-magnetic penguin operator}

In the effective Hamiltonian approach, the Hamiltonian is expressed as
the convolution of local operators and the Wilson coefficients. 
The effective Hamiltonian for $\Delta S = 1$ transitions is given by
\begin{eqnarray}
\hspace{-5mm}
H_{\rm eff}
&=&
\frac{G_F}{\sqrt{2}}
\left[
\sum_{q'=u,c}V_{q's}^* V_{q'b}
\left( C_1(\mu)O_1^{(q')}(\mu)
      + C_2(\mu)O_2^{(q')}(\mu)\right)
\right.
\nonumber\\
& &
\left.
- 
V_{ts}^* V_{tb}
\left(
\sum_{i=3}^{10}C_i(\mu)O_i(\mu)
+
C_{7\gamma}(\mu)O_{7\gamma}(\mu)
+
C_{8G}(\mu)O_{8G}(\mu)
\right) 
\right]
+ {\rm h.c.}
\;, 
\label{hbk}
\end{eqnarray}
where $V_{q's}^*$ and $V_{q'b}$ are the Cabibbo-Kobayashi-Maskawa
matrix elements,\cite{Cabibbo:yz,Kobayashi:fv} \
$O_{1-10}$ are local four-fermi operators, 
$O_{7\gamma}$ is the photo-magnetic penguin operator, and $O_{8G}$
is the chromo-magnetic penguin operator.
The local operators are given by
\begin{eqnarray}
& &O_1^{(q')} = (\bar{s}_iq'_j)_{V-A}(\bar{q'}_jb_i)_{V-A}\;,
\;\;\;\;\;\;\;\;
O_2^{(q')} = (\bar{s}_iq'_i)_{V-A}(\bar{q'}_jb_j)_{V-A}\;, 
\nonumber \\
& &O_3 =(\bar{s}_ib_i)_{V-A}\sum_{q}(\bar{q}_jq_j)_{V-A}\;,
\;\;\;\;\;\;\;
O_4 =(\bar{s}_ib_j)_{V-A}\sum_{q}(\bar{q}_jq_i)_{V-A}\;, 
\nonumber \\
& &O_5 =(\bar{s}_ib_i)_{V-A}\sum_{q}(\bar{q}_jq_j)_{V+A}\;,
\;\;\;\;\;\;\;
O_6 =(\bar{s}_ib_j)_{V-A}\sum_{q}(\bar{q}_jq_i)_{V+A}\;, 
\nonumber \\
& &O_7 =\frac{3}{2}(\bar{s}_ib_i)_{V-A}\sum_{q}e_q(\bar{q}_jq_j)_{V+A}\;,
\;
O_8 =\frac{3}{2}(\bar{s}_ib_j)_{V-A}\sum_{q}e_q(\bar{q}_jq_i)_{V+A}\;, 
\nonumber \\
& &O_9 =\frac{3}{2}(\bar{s}_ib_i)_{V-A}\sum_{q}e_q(\bar{q}_jq_j)_{V-A}\;,
\;
O_{10} =\frac{3}{2}(\bar{s}_ib_j)_{V-A}\sum_{q}e_q(\bar{q}_jq_i)_{V-A}\;, 
\nonumber \\
& &O_{7\gamma}  =  \frac{e}{8\pi^2} m_b \bar{s}_i \sigma^{\mu\nu}
          (1+\gamma_5) b_i F_{\mu\nu}\;,\;\;            
O_{8G}     =  -\frac{g_s}{8\pi^2} m_b \bar{s}_i \sigma^{\mu\nu}
   (1+\gamma_5)T^a_{ij} b_j G^a_{\mu\nu}\;,
\end{eqnarray} 
where $i$ and $j$ are color indices, and $q$ is taken to be
$u,\;d,\;s$ and $c$. We use the leading
logarithmic results of the Wilson  
coefficients in our calculations.\cite{Buchalla:1995vs} \  
We consider the chromo-magnetic penguin contribution for hadronic
two-body decays.
The non-local operator $O'_{8G}$ generated by $O_{8G}$ is given by
\begin{eqnarray}
O'_{8G}
 &=& - \frac{\alpha_s}{2\pi}m_b
   {\bar s}_i \sigma_{\mu\nu} T^a_{ij}(1+\gamma_5) b_j
\left[
q^\mu 
\frac{-i}{q^2} i
\left({\bar s}_{i'} \gamma^{\nu} T^a_{i'j'}s_{j'}\right)
(2 i)\right.
\nonumber\\
& &\left.\;\;
+ 4\pi \alpha_s f^{abc} 
\frac{-i}{q'^2} i
\left({\bar q}_{i'} \gamma^{\mu} T^b_{i'j'}q_{j'}\right)
\frac{-i}{q''^2} i
\left({\bar q}_{i''} \gamma^{\nu} T^c_{i''j''}q_{j''}\right)
\right. \nonumber\\
& &\left.\;\;
+ 
4\pi \alpha_s 
q^\mu 
f^{abc} 
\left[
   g^{\nu\lambda} (q + k_1)^\sigma
 + g^{\lambda\sigma} (-k_1+k_2)^\nu
 + g^{\sigma\nu} (-k_2-q)^\lambda
\right]
\frac{-i}{q^2} 
\right. \nonumber\\
& &\left.\;\;\;\;\;\;\;\;\;\;
\times 
\frac{-i}{k_1^2} i
\left(
  {\bar q}_{i'} \gamma_{\lambda} T^b_{i'j'} q_{j'}
\right)
\frac{-i}{k_2^2} i 
\left(
  {\bar q}_{i''} \gamma_{\sigma} T^c_{i''j''} q_{j''}
\right)
(2 i)
\right]   
\;,
\label{eq:chromo2}
\end{eqnarray}
where $q$ is the momentum transferred by the
gluon, which goes out from the $\sigma^{\mu\nu}$ vertex.
The second line in Eq.~(\ref{eq:chromo2}) is
induced by the self-interaction of gluons in $G^a_{\mu\nu}$, and the third
and fourth lines contain a 3-point vertex of gluons. The quantities 
$k_1$ and $k_2$ are the momenta that go out from the 3-point vertex of
gluons. These terms are needed to maintain gauge invariance.
The first term is first order in $\alpha_s$, and the others are of
order $\alpha_s^2$.
In the PQCD approach, all of the terms contribute to the same order in
$\alpha_s$, as we see later in the next section.

%-------------------------------------------------------------------%
%
%             Magnetic penguin amplitudes in PQCD approach
%
%-------------------------------------------------------------------%
\section{Magnetic penguin amplitudes in the PQCD approach}

\begin{figure}[hbt]
\begin{center}
\includegraphics{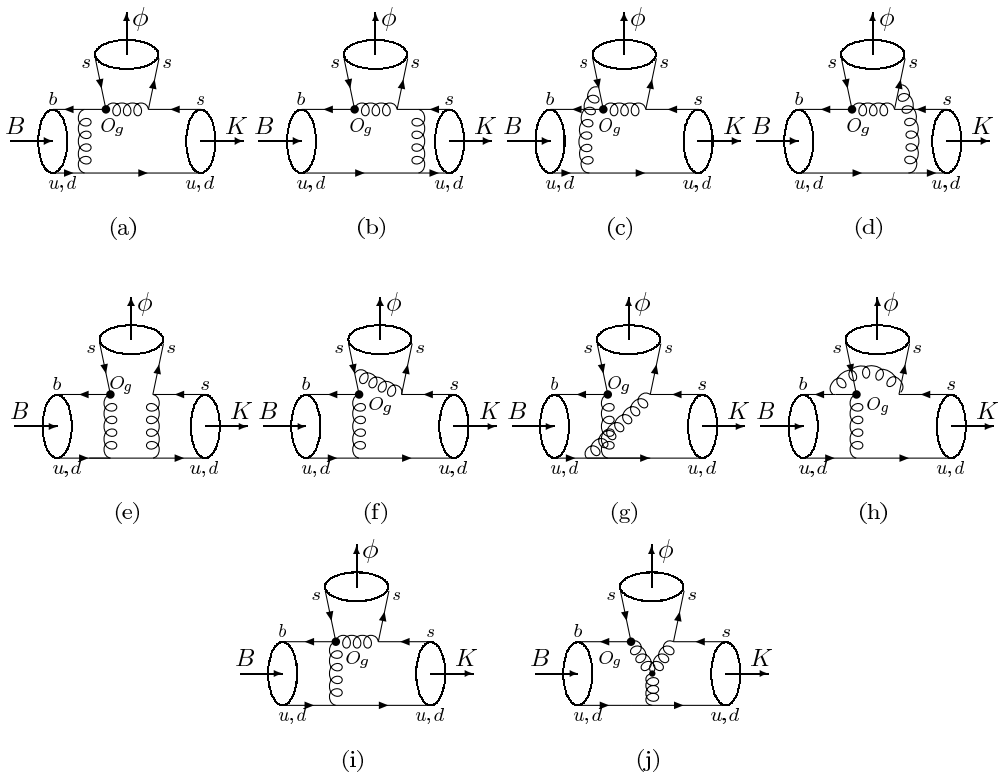}
\caption{Magnetic penguin diagrams in $B \to \phi K $ decays.}
\label{fig:magnetic}
\end{center}
\end{figure}

In this section, we calculate the chromo-magnetic penguin amplitudes in 
$B\to \phi K$ decays using the PQCD approach.
The chromo-magnetic penguin operator for hadronic two-body decays is
written as in Eq.~(\ref{eq:chromo2}), and there must be at least one hard
gluon emitted by the spectator quark in PQCD. 
Therefore, the diagrams with the chromo-magnetic penguin operator are
as shown in Fig.~\ref{fig:magnetic}.
The diagrams (a) -- (h) come from the first line in
Eq.~(\ref{eq:chromo2}), and the diagrams (i) and (j) come from the
other terms.
All of these diagrams contribute to the same order in $\alpha_s$.
The chromo-magnetic penguin contributions are of next-to-leading
order in $\alpha_s$ within the PQCD formalism. However, if we include
the Wilson coefficient, they are of the same order
in $\alpha_s$ as the penguin contributions. 
If we consider the full diagrams of the penguin and the chromo-magnetic 
penguin, we see that they are of same order in $\alpha_s$.
Therefore, we cannot say that the magnetic penguin contributions are
smaller than the penguin contributions.
In order to obtain a meaningful result for the branching ratios, we must
calculate other higher-order diagrams, which are the charm penguin, the
vertex corrections, and so on. 
In this study, we consider only the chromo-magnetic penguin, as this
operator is most sensitive to new physics.  

The decay width for $B \to \phi K $ decays is written as
\begin{equation}
\Gamma=\frac{G_F^2}{32\pi M_B}|{\cal A}|^2\;,
\end{equation}
where ${\cal A}$ is the sum of the leading amplitude and the magnetic
penguin amplitude. The decay amplitudes from the magnetic penguin are
given by
\begin{eqnarray}
{\cal A}^{MP} 
&=& 
V^*_{tb}V_{ts} \left( 
  {\cal M}^{MP}_a + {\cal M}^{MP}_b
+ {\cal M}^{MP}_c + {\cal M}^{MP}_d
+ {\cal M}^{MP}_e + {\cal M}^{MP}_f
\right. \nonumber\\
& & \left.\hspace{20mm}\;
+ {\cal M}^{MP}_g + {\cal M}^{MP}_h
+ {\cal M}^{MP}_i + {\cal M}^{MP}_j
\right) \;,
\end{eqnarray}
where the indices denote the labels in Fig.~\ref{fig:magnetic}.
The charged modes and neutral modes in $B\to\phi K$ have the same
contributions of the chromo-magnetic penguin.
The amplitudes ${\cal M}$ are expressed as the convolutions of meson wave 
functions, a hard part, and the Wilson coefficients.\cite{Chang:1996dw} \ 
The meson wave functions are non-perturbative, and the hard part, which
includes the exchange of a hard gluon, is perturbative.
We introduce small transverse momenta for quarks and anti-quarks in
mesons. Large double logarithms are then generated by corrections in the 
meson wave functions. Their resummation leads to the Sudakov
factor.\cite{Botts:kf} \  
The Sudakov factor guarantees a perturbative calculation of the hard
part.\cite{Li:1992nu} \ 
The other double logarithms appear from the end-point region of the
parton momenta. The resummation of these logarithms leads to the
threshold factor in the hard part.\cite{Li:2001ay} \ 

We consider the $B$ meson to be at rest.
In the light-cone coordinates, the $B$ meson momentum $P_1$, the $K$
meson momentum $P_2$, and the $\phi$ meson momentum $P_3$ are taken to be
\begin{eqnarray}
P_1=\frac{M_B}{{\sqrt 2}}(1,1,{\bf 0}_T)\;,\;\;
P_2=\frac{M_B}{{\sqrt 2}}(1-r_\phi^2,0,{\bf 0}_T)\;,\;\;
P_3=\frac{M_B}{{\sqrt 2}}(r_\phi^2,1,{\bf 0}_T)\;,
\end{eqnarray}
where $r_\phi = M_\phi/M_B$, and the $K$ meson mass is ignored.  
The momentum of the spectator quark in the $B$ meson is written as $k_1$.
The light quark in the $B$ meson has a minus component $k_1^-$ given by
$k_1^-=x_1P_1^-$, where $x_1$ is the momentum fraction. 
The quarks in the $K$ meson have the plus components $x_2P_2^+$ and
$(1-x_2)P_2^+$ and the small transverse components $\boldsymbol{k}_{2T}$ and
$-\boldsymbol{k}_{2T}$, respectively. 
The quarks in the $\phi$ meson have the minus components $x_3P_3^-$ and
$(1-x_3)P_3^-$ and the small transverse components $\boldsymbol{k}_{3T}$ and
$-\boldsymbol{k}_{3T}$, respectively. 
The $\phi$ meson longitudinal polarization vector $\epsilon_\phi$ and
two transverse polarization vectors $\epsilon_{\phi T}$ are given by 
$\epsilon_\phi=(1 / \sqrt{2}r_\phi)( -r_\phi^2, 1, \boldsymbol{0}_T)$
and 
$\epsilon_{\phi T}=(0,0,\boldsymbol{1}_T)$.

For example, the amplitude for Fig.~\ref{fig:magnetic}(a) is written as
\begin{eqnarray}
{\cal M}^{MP}_a
&=&
- 8 M_B^6 C_F^2 \frac{\sqrt{2N_c}}{2N_c}
\int_0^1 dx_1 dx_2 dx_3\int_0^{\infty} b_1db_1 b_2db_2 b_3db_3 
\phi_B(x_1,b_1)
\nonumber \\
& & \times
\left[ 
    - (1-x_2)
      \left\{ 2 \phi_K^A(x_2)
        + r_K \left( 3 \phi_K^P(x_2)+\phi_K^T(x_2) \right)
    \right. \right.\nonumber \\
    & & \;\;\;\;\;\;\; \;\;\;\;\;\;\; \;\;\;\;\;\;\; \left.\left.
        + r_K x_2 \left( \phi_K^P(x_2) - \phi_K^T(x_2) \right) 
    \right\} \phi_{\phi}(x_3)
    \right. \nonumber \\
    & & \;\;\;\;\;\;\; \left.
    - r_\phi (1+x_2) x_3 \phi_K^A(x_2) 
      \left( 3 \phi_{\phi}^s(x_3) - \phi_{\phi}^t(x_3) \right) 
    \right. \nonumber \\
    & & \;\;\;\;\;\;\; \left.
    - r_\phi r_K(1-x_2)
        \left( \phi_K^P(x_2) - \phi_K^T(x_2) \right)
        \left( 3 \phi_{\phi}^s(x_3) + \phi_{\phi}^t(x_3) \right) 
    \right. \nonumber \\
    & & \;\;\;\;\;\;\; \left.
    - r_\phi r_K x_3(1-2x_2) 
        \left( \phi_K^P(x_2) + \phi_K^T(x_2) \right)
        \left( 3 \phi_{\phi}^s(x_3) - \phi_{\phi}^t(x_3) \right) 
\right]
\nonumber \\
& & \times
E_{g}(t) N_t \{x_2(1-x_2)\}^c h_{e}^{MP}(A,B,C,b_1,b_2,b_3)
\;,
\label{eq:ma}
\end{eqnarray}
where $ r_K \equiv m_{0K}/M_B$ and $m_{0K}$ is the chiral factor defined as
$m_{0K}\equiv M_K^2/(m_d+m_s)$.
The meson distribution amplitudes, $\phi_B$, $\phi_K^A$, and so on, are
given in the Appendix~\ref{ap:WF}. 
$N_t \{x_2(1-x_2)\}^c$ is the threshold factor, where $N_t=1.775$ and
$c=0.3$.\cite{Kurimoto:2001zj} \  
The evolution factors $E_{g}(t)$ are defined by 
$E_{g}(t)=\{\alpha_s(t)\}^2 C_{8G}(t)\exp[-S_B(t)-S_K(t)-S_\phi(t)]$, in
which $S_B(t)$, $S_K(t)$ and $S_\phi(t)$ are the Sudakov factors given
in the Appendix~\ref{ap:SF}.
The hard functions are given by
\begin{eqnarray}
h_e^{MP}(A,B,C,b_1,b_2,b_3)
&=&
- K_{0}\left( B b_1 \right)
  K_0  \left( C b_3 \right)
\nonumber\\
& &\hspace{-15mm}
\times
\int_0^{\frac{\pi}{2}} d\theta
  \tan\theta
  J_0\left( A b_1 \tan\theta \right)
  J_0\left( A b_2 \tan\theta \right)
  J_0\left( A b_3 \tan\theta \right)
\;,
\label{eq:he}
\end{eqnarray}
where $A$, $B$ and $C$ are given on Table~\ref{tab:ABC2}.
The values $A^2$, $B^2$ and $C^2$ are the squares of the virtual quark
and virtual gluons momenta.
We assume that the hard scale $t$ is defined as
\begin{eqnarray}
t = {\rm max}(\sqrt{|A^2|}, \sqrt{|B^2|}, 1/b_1, 1/b_2, 1/b_3)
\;.
\label{eq:hs}
\end{eqnarray}
In the above definition of $t$, we have chosen the same definition for $t$ as
we employed for computation of the leading diagrams. 
There is a slight dependence on $t$ in the chromo-magnetic penguin
amplitude, as we see in the next section.
The expressions for other diagrams are summarized in the
Appendix~\ref{ap:MPA}.

%-------------------------------------------------------------------%
%
%                Numerical Result
%
%-------------------------------------------------------------------%
\section{Numerical result}

The parameters that we used in this calculation are as
follows:\cite{Hagiwara:fs} \ 
$M_B = 5.28\; {\rm GeV}$, 
$M_K =0.49 \;{\rm GeV}$, 
$M_{\phi}=1.02\; {\rm GeV}$,
$m_b=4.8 \; {\rm GeV}$, 
$m_t = 174.3 \; {\rm GeV}$,
$f_{B}= 190\; {\rm MeV}$, 
$f_{K} = 160\; {\rm MeV}$, 
$f_{\phi} = 237 \;{\rm MeV}$,
$f_{\phi}^T = 220 \;{\rm MeV}$,
$\tau_{B^0}=1.54\times 10^{-12}\;{\rm sec}$, 
$\tau_{B^\pm}=1.67\times 10^{-12} \;{\rm sec}$, 
$\Lambda_{\rm QCD}^{(4)}=0.250\;{\rm GeV}$.  
In addition, we used the value of the chiral factor $m_{0K} =1.70$ GeV, 
and the parameter values in wave functions are as given in the
Appendix~\ref{ap:WF}. 

The numerical results are listed in Table~\ref{tab:phik_mag}.
%
%%%%%%%%%%%%%%%%%%%%%%%%%%%%%%%%%%%%%%%%%%%%%%%%%%%%%%%%%%%%%%%%%%
\begin{table}[hbt]
\caption{Magnetic penguin contributions for $B\to\phi K$.}    
\label{tab:phik_mag}
\begin{center}
\begin{tabular}{cc}
\hline\hline
${\cal M}^{MP}_a$ + ${\cal M}^{MP}_b$ &
$ -0.0308\; - i\; 0.0169 $ 
\\
${\cal M}^{MP}_e$ + ${\cal M}^{MP}_f$ &
$\;\;\; 0.00309 \; + i\; 0.00544 $
\\
${\cal M}^{MP}_c$ + ${\cal M}^{MP}_d$ 
& \hspace{4mm}negligible 
\\
+ ${\cal M}^{MP}_g$ + ${\cal M}^{MP}_h$
& \\
${\cal M}^{MP}_i$ &
$\;\; -0.00229\; - i\; 0.000798 $  
\\
${\cal M}^{MP}_j$ &
$\; 0.00623 \; - i\; 0.0147 $
\\
\hline
${\cal M}^{MP}$  &$ -0.0238\; - i\; 0.0270 $
\\
\hline
\end{tabular}
\end{center}
\end{table}
%%%%%%%%%%%%%%%%%%%%%%%%%%%%%%%%%%%%%%%%%%%%%%%%%%%%%%%%%%%%%%%%%%
%
The leading amplitude for $B^0\to\phi K^0$ is
$0.0944-i\;0.0383$.\cite{Mishima:2001pp} \  
We find that the diagrams (a) and (b) are
dominant, and the magnetic penguin amplitudes are about 30\% of the
leading amplitudes.

\begin{figure}[htbp]
\begin{minipage}[t]{0.45\textwidth}
\begin{center}
\includegraphics{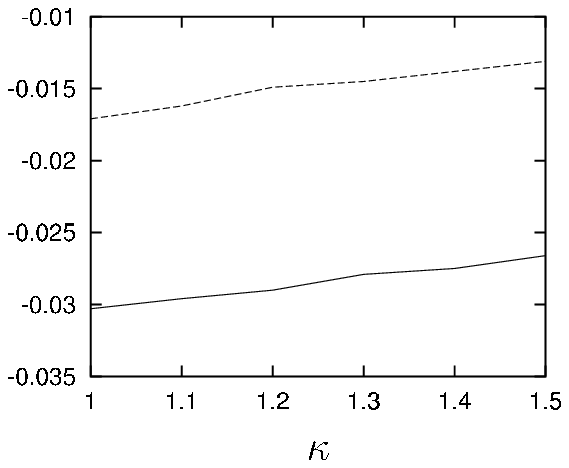}
\caption{$t$ dependence of ${\cal M}^{MP}_a+{\cal M}^{MP}_b$. The lower
 curve is the real part and the upper curve is the imaginary part. We
 see that there
 is a slight dependence on $t$ in the chromo-magnetic penguin.}
\label{fig:t}
\end{center}
\end{minipage}
\begin{minipage}[t]{0.04\textwidth}
\hspace{1mm}
\end{minipage}
\begin{minipage}[t]{0.45\textwidth}
\begin{center}
\includegraphics{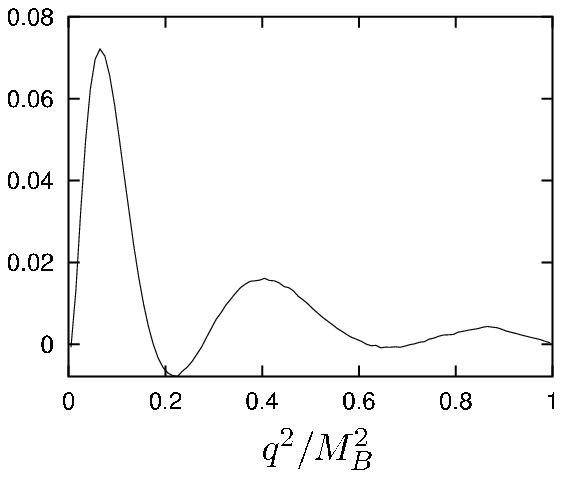}
\caption{The distribution of $q^2$ for ${\rm Re}{\cal M}^{MP}_a$. The
 expectation value of $q^2$ is 6.3 GeV$^2$.}
\label{fig:q2}
\end{center}
\end{minipage}
\end{figure}

We study the hard scale dependence of chromo-magnetic amplitudes.
We define the hard scale as $t=\kappa \cdot {\rm max}(\sqrt{|A^2|},
\sqrt{|B^2|}, 1/b_1, 1/b_2, 1/b_3)$, where $\kappa$ is a parameter, and
vary $\kappa$ between 1 and 1.5. 
Figure~\ref{fig:t} shows the hard scale dependence of ${\cal
M}^{MP}_a+{\cal M}^{MP}_b$. We find that  
there is a slight dependence on $t$ in the chromo-magnetic penguin.
This is consistent with the fact that the Wilson coefficient for
$O_{8G}$ is most sensitive to new physics, and its contribution, by
itself, is physical. 

The value of $q^2$ of the virtual gluon in $O'_{8G}$ is written in terms
of $x_1$, 
$x_2$ and $x_3$, which are the momentum fractions of partons.
For example, in the diagram (a), $q^2 = (1-x_2)x_3M_B^2$. In
Fig.~\ref{fig:q2}, we show the distribution of $q^2$ for 
${\cal M}^{MP}_a$. The shape of this graph is not simple. 
The expectation value of $q^2$ is calculated as
\begin{equation}
\langle q^2 \rangle = 6.3\;\; {\rm GeV}^2\;.
\label{eq:q2}
\end{equation}
This is approximately equal to $M_B^2/4$.

Next, we analyze the FA using the value of $q^2$ given in Eq.~(\ref{eq:q2}).
In the FA, the effective Hamiltonian for the chromo-magnetic penguin
is\cite{Ali:1997nh} \   
\begin{eqnarray}
H_{\rm eff}
=
- \frac{G_F}{\sqrt{2}} V_{ts}^* V_{tb} \frac{\alpha_s}{8\pi} 
\left( -
\frac{2m_b}{\sqrt{\langle q^2 \rangle}} C_{8G}
\right)
\left(
O_4+O_6-\frac{1}{N_c}\left(O_3+O_5\right)
\right)
\;.
\end{eqnarray}
We use the value $6.3$ GeV$^2$ for $\langle q^2 \rangle$ and $1.5$ GeV
for the renormalization scale, which is a typical scale of the hard part
in the PQCD approach.  
The result for the amplitude is ${\cal M} = - 0.029$, where we input the
value $F_{BK}=0.38$. 
This is to be compared with ${\cal M}^{MP}_a + {\cal M}^{MP}_b = -0.0308
-i\;0.0169$. The real part is agree with the above result.

%-------------------------------------------------------------------%
%
%                Summary
%
%-------------------------------------------------------------------%
\section{Summary}

In this paper, we calculated the chromo-magnetic penguin amplitudes for
$B\to\phi K$ decays in the PQCD approach.
We find that the chromo-magnetic penguin contribution is about 30\% of
the leading contribution. Therefore, we find that the chromo-magnetic
penguin is not 
negligible in hadronic two-body decays. However, for the Standard Model
there are many other 
higher-order diagrams that must be considered simultaneously if we are
to add the magnetic penguin term. 
If we consider the contribution of new physics, this diagram can be
considered separately. It is noted that there is only weak energy scale
dependence.   
In addition, we calculated the expectation value of $q^2$
of the gluon in the chromo-magnetic penguin, and the result is
approximately $M_B^2/4$.

%-------------------------------------------------------------------%
%
%                 Acknowledgements
%
%-------------------------------------------------------------------%
\section*{Acknowledgements}

We would like to thank the PQCD group members: C.~H.~Chen, Y.Y.~Keum,
E.~Kou, T.~Kurimoto, H-n.~Li, C.~D.~Lu, T.~Morozumi and K.~Ukai 
for useful discussions. 
S.~M. was supported by the Japan Society for the Promotion of Science
under the Predoctoral Research Program.
A.~I.~S. acknowledges support from the Japan Society for the Promotion
of Science Japan-US collaboration program and a grant from the Ministry of
Education, Culture, Sports, Science and Technology of Japan.

\newpage
%-------------------------------------------------------------------%
%
%                  Appendix
%
%-------------------------------------------------------------------%
\appendix

\section{The Sudakov Factor}\label{ap:SF}

The Sudakov factor is written as\cite{Li:1992nu} \ 
\begin{eqnarray}
S_B(t)&=&s(x_1P_1^+,b_1) + \int_{1/b_1}^t 
        \frac{d\bar\mu}{\bar\mu}\gamma_\phi(\alpha_s(\bar\mu))
\;,
\\
S_{K}(t)&=&s(x_2P_2^+,b_2)+s((1-x_2)P_2^+,b_2) + \int_{1/b_2}^t 
        \frac{d\bar\mu}{\bar\mu}\gamma_\phi(\alpha_s(\bar\mu))
\;,
\\
S_\phi(t)&=&s(x_3P_3^-,b_3)+s((1-x_3)P_3^-,b_3) + \int_{1/b_3}^t 
        \frac{d\bar\mu}{\bar\mu}\gamma_\phi(\alpha_s(\bar\mu))
\;.
\end{eqnarray}
The exponent $s$ is given by
\begin{eqnarray}
s(Q,b)=\int_{1/b}^{Q}\frac{d \mu}{\mu}\left[\ln\left(\frac{Q}{\mu}
\right)A(\alpha_s(\mu))+B(\alpha_s(\mu))\right]\;,
\end{eqnarray}
where the anomalous dimensions $A$, to two loops, and $B$, to one loop, are
\begin{eqnarray}
A&=&C_F\frac{\alpha_s}{\pi}+\left[\frac{67}{9}-\frac{\pi^2}{3}
-\frac{10}{27}f+\frac{2}{3}\beta_0\ln\left(\frac{e^{\gamma_E}}{2}\right)
\right]\left(\frac{\alpha_s}{\pi}\right)^2\;,
\\
B&=&\frac{2}{3}\frac{\alpha_s}{\pi}\ln\left(\frac{e^{2\gamma_E-1}}
{2}\right)\;,
\end{eqnarray}
with $C_F=4/3$ being a color factor, $f$ being the number of active
flavors, $\gamma_E$ being the Euler constant, and $\beta_{0}=(33-2f)/3$.
The anomalous dimension of mesons is given by
\begin{eqnarray}
\gamma_\phi(\alpha_s(\mu))
= 2 \gamma_q(\alpha_s(\mu)) = -2\frac{\alpha_s(\mu)}{\pi}\;.
\end{eqnarray}

\section{Wave Functions}\label{ap:WF}

The $B$ meson wave function is defined by
\begin{eqnarray}
\Phi^{({\rm in})}_{B,\alpha\beta,ij}
&\equiv& 
\langle 0 |{\bar b}_{\beta j}(0)d_{\alpha i}(z) | B(P) \rangle
\nonumber\\
&=&
\frac{i \delta_{ij}}{\sqrt{2N_c}}\int dx d^2{\boldsymbol{k}_{T}}
 e^{-i(xP^-z^+-{\boldsymbol{k}_{T}}{\boldsymbol{z}_T})}
\left[
(\not P +M_B)\gamma_5
 \phi_B(x,\boldsymbol{k}_{T})
\right]_{\alpha\beta}
\;,
\end{eqnarray}
where the indices $\alpha$ and $\beta$ are spin indices, $i$ and $j$ are
color indices, and $N_c$ is the color factor. The distribution amplitude
$\phi_B$ is normalized as 
\begin{eqnarray}
\int_0^1 dx_{1}\phi_B(x_{1},b_1=0)&=&\frac{f_B}{2\sqrt{2N_c}}\;,
\label{eq:bnor}
\end{eqnarray}
where $b_1$ is the conjugate space of $k_1$, and $f_B$ is the decay
constant of the $B$ meson. In this study, we used the model functions  
\begin{eqnarray}
\phi_B(x,b) &=& N_B
x^2 (1-x)^2\exp\left[-\frac{1}{2}\left(\frac{xM_B}{\omega_{B}}\right)^2
-\frac{\omega_{B}^2 b^2}{2}\right] \;,
\end{eqnarray}
where $N_B$ is the normalization constant and $\omega_B$ is the shape
parameter. We used $\omega_B=0.4$ GeV, which was determined by the
calculation of form factors.\cite{Kurimoto:2001zj} \  

The $K$ meson wave function is given by
\begin{eqnarray}
\Phi^{({\rm out})}_{K,\alpha\beta,ij}
&\equiv&
\langle K(P) |\bar d_{\beta j}(z) s_{\alpha i}(0)|0\rangle
\nonumber\\
& & \hspace{-20mm} =
\frac{-i\delta_{ij}}{\sqrt{2N_c}}\int^1_0dx e^{ixP\cdot z}
\gamma_5\left[
\not P
 \phi^A_K(x)+m_{0K}
 \phi_K^P(x)
+m_{0K} (\not v\not n -1)
 \phi_K^T(x)
\right]_{\alpha\beta}
,
\end{eqnarray}
where $m_{0K} = M_K^2/(m_d+m_s)$, $n_\mu\equiv z_\mu/z^-$ and
$v_\mu\equiv P_{\mu}/P^+$. The function $\phi^A_K(x)$ is the leading-twist
contribution and $\phi_K^P$ and $\phi_K^T$ are twist-3. These
distribution amplitudes were calculated using the light-cone QCD sum
rule:\cite{Ball:1998tj} \
\begin{eqnarray}
\phi_{K}^A(x) &=& \frac{f_K}{2\sqrt{2N_c}}6x(1-x)
\left[1 + a_1 C_1^{\frac{3}{2}}(1-2x)
+ a_2C_2^{\frac{3}{2}}(1-2x)\right]
\;,\\
\phi^P_{K}(x) &=& \frac{f_K}{2\sqrt{2N_c}}
\bigg[ 
1
+\left(30\eta_3 -\frac{5}{2}\rho_K^2\right)
C_2^{\frac{1}{2}}(1-2x)
\nonumber\\
& &\;\;\;\;\;\;\;\;\;\;\;\;\;\;\;\;\;\;\;\;
- 3\left\{
\eta_3\omega_3 + \frac{9}{20}\rho_K^2(1+6a_2)   
\right\}
C_4^{\frac{1}{2}}(1-2x) 
\bigg]
\;,\\
\phi^T_{K}(x) &=& \frac{f_K}{2\sqrt{2N_c}}(1-2x)
\nonumber\\
& &\;\;\; \times
\bigg[ 1
+
6\left(5\eta_3 -\frac{1}{2}\eta_3\omega_3 - \frac{7}{20}
      \rho_K^2 - \frac{3}{5}\rho_K^2 a_2 \right)
(1-10x+10x^2) \bigg]
\;.
\end{eqnarray}
Here, $\rho_K=(m_d+m_s)/M_K$ and $C_n^{\nu}(x)$ is the Gegenbauer
 polynomial. The parameter values are $a_1=0.17$, $a_2=0.20$,
 $\eta_3=0.015$ and $\omega_3=-3.0$.

The $\phi$ meson wave function of the longitudinal parts is given by
\begin{eqnarray}
\Phi^{({\rm out})}_{\phi,\alpha\beta,ij}
&\equiv& 
\langle \phi(P_3) | \bar s_{\beta j}(z) s_{\alpha i}(0) | 0 \rangle
\nonumber\\
& & \hspace{-20mm} =
\frac{\delta_{ij}}{\sqrt{2N_c}}\int^1_0dx_3 e^{ix_3P_3\cdot z}
\left[
M_\phi\not\epsilon_\phi 
 \phi_\phi(x_3)+
\not\epsilon_\phi \not P_3
 \phi_\phi^t(x_3)
+ M_\phi
 \phi_\phi^s(x_3)
\right]_{\alpha\beta}
\;.
\end{eqnarray}
Here, $\phi_\phi$ is the leading-twist distribution amplitude, and the other
terms are the twist-3 amplitudes. They were also calculated using the
light-cone QCD sum rule:\cite{Ball:1998sk} \  
\begin{eqnarray}
\phi_{\phi}(x) &=& \frac{f_{\phi}}{2\sqrt{2N_c}}6x(1-x)
\;,\\
\phi^t_{\phi}(x) &=& \frac{f^T_{\phi}}{2\sqrt{2N_c}}
\bigg[ 3(1-2x)^2
+\frac{35}{4}\zeta_3^T\{3-30(1-2x)^2+35(1-2x)^4\} 
\nonumber\\
& &\;\;\;\;\;\;\;\;\;\;\;\;\;\;\;\;\;\;\;\;
+\frac{3}{2}\delta_{+}\left\{1-(1-2x)\log\frac{1-x}{x}\right\} \bigg]
\;,\\
\phi^s_{\phi}(x) &=& \frac{f^T_{\phi}}{4\sqrt{2N_c}}
\bigg[ 
(1-2x)\left\{
6+9\delta_{+}+140\zeta_3^T (1-10x+10x^2)
\right\}
\nonumber\\
& &\;\;\;\;\;\;\;\;\;\;\;\;\;\;\;\;\;\;\;\;
+3\delta_{+}\log\frac{x}{1-x}
 \bigg]\;.
\end{eqnarray}
Here, $\zeta_3^T=0.024$ and $\delta_{+}=0.46$.

\section{Magnetic Penguin Amplitudes}\label{ap:MPA}

Below, we present the decay amplitudes for Fig.~\ref{fig:magnetic} (b)
-- (j): 
\begin{eqnarray}
{\cal M}^{MP}_b
&=&
8 M_B^6 C_F^2 \frac{\sqrt{2N_c}}{2N_c}
\int_0^1 dx_1 dx_2 dx_3\int_0^{\infty} b_1db_1 b_2db_2 b_3db_3 
\phi_B(x_1,b_1)
\nonumber \\
& & \times
\left[ 
    x_1 \phi_K^A(x_2)\phi_{\phi}(x_3) -2 r_K (-2+x_1) \phi_K^P(x_2)
         \phi_{\phi}(x_3)
    \right. \nonumber \\
    & & \;\;\;\;\;\;\; \left.
    + r_\phi x_1 \phi_K^A(x_2)
         \left( 3 \phi_{\phi}^s(x_3) + \phi_{\phi}^t(x_3) \right) 
    \right. \nonumber \\
    & & \;\;\;\;\;\;\; \left.
    - 12 r_\phi r_K x_1 
        \phi_K^P(x_2) \phi_{\phi}^s(x_3)
    + 2 r_\phi r_K x_3 \phi_K^P(x_2) 
         \left( 3 \phi_{\phi}^s(x_3) - \phi_{\phi}^t(x_3) \right) 
\right]
\nonumber \\
& & \times 
E_{g}(t) N_t \{x_1(1-x_1)\}^c h_e^{MP}(A,B,C,b_2,b_1,b_3)
\;,\\
%
%------------------------------------------------------------------
%
{\cal M}^{MP}_c 
&=& 
8 M_B^6 C_F \left(C_F-\frac{N_c}{2}\right) \frac{\sqrt{2N_c}}{2N_c}
\int_0^1 dx_1 dx_2 dx_3\int_0^{\infty} b_1db_1 b_2db_2 b_3db_3 
\phi_B(x_1,b_1)
\nonumber \\
& & \times
\left[
    2 (1-x_2)(1-x_1-x_3) \phi_K^A(x_2) \phi_{\phi}(x_3)
      \right. \nonumber \\
      & & \;\;\;\;\;\;\;\; \left.
    + r_K (1-x_2)(1-x_1-x_2-x_3) 
        \left( \phi_K^P(x_2) - \phi_K^T(x_2) \right) \phi_{\phi}(x_3) 
      \right. \nonumber \\
      & & \;\;\;\;\;\;\;\; \left.
    + r_\phi x_3(1-x_1-x_2-x_3) \phi_K^A(x_2) 
        \left( \phi_{\phi}^s(x_3) + \phi_{\phi}^t(x_3) \right) 
      \right. \nonumber \\
      & & \;\;\;\;\;\;\;\; \left.
    + r_K r_\phi \left\{ (1-x_1)(1-x_2) - x_3 \right\}
        \left( \phi_K^P(x_2) - \phi_K^T(x_2) \right)
        \left( 3\phi_{\phi}^s(x_3) + \phi_{\phi}^t(x_3) \right) 
      \right. \nonumber \\
      & & \;\;\;\;\;\;\;\; \left.
    + 2 r_K r_\phi x_2 x_3 
        \left( 3 \phi_K^P(x_2)\phi_{\phi}^s(x_3) 
                    - \phi_K^T(x_2)\phi_{\phi}^t(x_3) \right) 
\right]
\nonumber \\
& & \times 
E_{g}(t) h_n^{MP}(A,B,C,b_1,b_2,b_3)
\;,\\
%
%------------------------------------------------------------------
%
{\cal M}^{MP}_d
&=&
- 8 M_B^6 C_F \left(C_F-\frac{N_c}{2}\right) \frac{\sqrt{2N_c}}{2N_c}
\int_0^1 dx_1 dx_2 dx_3\int_0^{\infty} b_1db_1 b_2db_2 b_3db_3 
\phi_B(x_1,b_1)
\nonumber \\
& & \times (x_1-x_3)
\left[ 
    -  (2+x_2)  
         \phi_K^A(x_2) \phi_{\phi}(x_3)
    \right.\nonumber \\
    & & \;\;\;\;\;\;\;\;\left. 
    - r_K \left\{ 
                  \left(\phi_K^P(x_2)-\phi_K^T(x_2)\right) 
               - x_2
                     \left(\phi_K^P(x_2)+\phi_K^T(x_2)\right) 
          \right\} \phi_{\phi}(x_3)
    \right.\nonumber \\
    & & \;\;\;\;\;\;\;\;\left. 
    + r_\phi \left( 2x_1-x_2-2x_3 \right) 
          \phi_K^A(x_2) 
          \left( \phi_{\phi}^s(x_3) - \phi_{\phi}^t(x_3) \right)
    \right.\nonumber \\
    & & \;\;\;\;\;\;\;\;\left. 
    + 2 r_K r_\phi 
          \left\{ 
                     \left(\phi_K^P(x_2)-\phi_K^T(x_2)\right) 
                  - x_2  
                     \left(\phi_K^P(x_2)+\phi_K^T(x_2)\right) 
          \phi_{\phi}^t(x_3)
          \right\}
\right]
\nonumber \\
& & \times
E_{g}(t) h_n^{MP}(A,B,C,b_2,b_1,b_3)
\;,
\\
%
%------------------------------------------------------------------
%
{\cal M}^{MP}_e
&=& 
8 M_B^6 C_F^2 \frac{\sqrt{2N_c}}{2N_c}
\int_0^1 dx_1 dx_2 dx_3\int_0^{\infty} b_1db_1 b_2db_2 b_3db_3 
\phi_B(x_1,b_1)
\nonumber \\
& & \times
\left[
    3 x_3 \phi_K^A(x_2) \phi_{\phi}(x_3)
    + 2 r_K \left( 2-x_1+x_3 \right) \phi_K^P(x_2) \phi_{\phi}(x_3)
    \right.\nonumber \\
    & & \;\;\;\;\;\;\;\; \left.
    + r_\phi x_3 \left\{ 1-2\left(x_1-x_3\right) \right\} \phi_K^A(x_2) 
            \left(\phi_{\phi}^s(x_3)-\phi_{\phi}^t(x_3)\right)
    \right.\nonumber \\
    & & \;\;\;\;\;\;\;\; \left.
    - 6 r_K r_\phi x_1 \phi_K^P(x_2)
            \left(\phi_{\phi}^s(x_3)-\phi_{\phi}^t(x_3)\right)
    + 12 r_K r_\phi x_3  \phi_K^P(x_2) \phi_{\phi}^s(x_3)
\right]
\nonumber \\
& & \times 
E_{g}(t) N_t \{x_3(1-x_3)\}^c h_e^{MP}(A,B,C,b_2,b_3,b_1)
\;,\\
%
%------------------------------------------------------------------
%
{\cal M}^{MP}_f
&=&
- 8 M_B^6 C_F^2 \frac{\sqrt{2N_c}}{2N_c}
\int_0^1 dx_1 dx_2 dx_3\int_0^{\infty} b_1db_1 b_2db_2 b_3db_3 
\phi_B(x_1,b_1)
\nonumber \\
& & \times
\left[
    - 3x_1(1-x_2) \phi_K^A(x_2) \phi_{\phi}(x_3)
    \right. \nonumber \\
    & & \;\;\;\;\;\;\; \left.
    + r_K (1-x_2)(2x_1-x_2) 
          \left( \phi_K^P(x_2) + \phi_K^T(x_2) \right) \phi_{\phi}(x_3)
      \right. \nonumber \\
      & & \;\;\;\;\;\;\;\; \left.
    + 2 r_\phi (x_1-2x_2) \phi_K^A(x_2) \phi_{\phi}^s(x_3)
    \right. \nonumber \\
    & & \;\;\;\;\;\;\; \left.
    - 6 r_Kr_\phi x_1(1-x_2)
          \left( \phi_K^P(x_2) - \phi_K^T(x_2) \right) \phi_{\phi}^s(x_3)
      \right. \nonumber \\
      & & \;\;\;\;\;\;\;\; \left.
    + 6 r_Kr_\phi x_2 
          \left( \phi_K^P(x_2) + \phi_K^T(x_2) \right )\phi_{\phi}^s(x_3)
\right]
\nonumber \\
& & \times 
E_{g}(t) N_t \{x_2(1-x_2)\}^c h_e^{MP}(A,B,C,b_3,b_2,b_1)
\;,\\
%
%------------------------------------------------------------------
%
{\cal M}^{MP}_g
&=& 
8 M_B^6 C_F \left(C_F-\frac{N_c}{2}\right) \frac{\sqrt{2N_c}}{2N_c}
\int_0^1 dx_1 dx_2 dx_3\int_0^{\infty} b_1db_1 b_2db_2 b_3db_3 
\phi_B(x_1,b_1)
\nonumber \\
& & \times 
(x_1-x_3)
\left[
    - (1-x_2) \phi_K^A(x_2) \phi_{\phi}(x_3)
    - r_K x_2 \left( \phi_K^P(x_2) - \phi_K^T(x_2) \right) 
          \phi_{\phi}(x_3)
      \right. \nonumber \\
      & & \;\;\;\;\;\;\;\; \left.
    - 2 r_K \phi_K^T(x_2) \phi_{\phi}(x_3)
    + r_\phi (1+x_2)\phi_K^A(x_2) 
          \left(\phi_{\phi}^s(x_3) - \phi_{\phi}^t(x_3)\right) 
    \right. \nonumber \\
    & & \;\;\;\;\;\;\; \left.
    - 4 r_\phi \phi_K^A(x_2) \phi_{\phi}^s(x_3)
      \right. \nonumber \\
      & & \;\;\;\;\;\;\;\; \left.
    + 2 r_K r_\phi x_2 
          \left( \phi_K^P(x_2) - \phi_K^T(x_2) \right) \phi_{\phi}^t(x_3) 
    + 4 r_Kr_\phi \phi_K^T(x_2) \phi_{\phi}^t(x_3) 
\right]
\nonumber \\
& & \times 
E_{g}(t) h_n^{MP}(A,B,C,b_2,b_3,b_1)
\;,\\
%
%------------------------------------------------------------------
%
{\cal M}^{MP}_h
&=&
8 M_B^6 C_F \left(C_F-\frac{N_c}{2}\right) \frac{\sqrt{2N_c}}{2N_c}
\int_0^1 dx_1 dx_2 dx_3\int_0^{\infty} b_1db_1 b_2db_2 b_3db_3 
\phi_B(x_1,b_1)
\nonumber \\
& & \times
\left[
    - x_1(1-x_2) \phi_K^A(x_2) \phi_{\phi}(x_3)
    - r_K x_2(1-x_2)
          \left( \phi_K^P(x_2) + \phi_K^T(x_2) \right) \phi_{\phi}(x_3) 
      \right. \nonumber \\
      & & \;\;\;\;\;\;\;\; \left.
    - r_K x_2(2-x_1)
          \left( \phi_K^P(x_2) - \phi_K^T(x_2) \right) \phi_{\phi}(x_3) 
    \right. \nonumber \\
    & & \;\;\;\;\;\;\; \left.
    + r_\phi x_1(x_1+x_3)  \phi_K^A(x_2)
          \left( \phi_{\phi}^s(x_3) - \phi_{\phi}^t(x_3) \right)
      \right. \nonumber \\
      & & \;\;\;\;\;\;\;\; \left.
    + r_\phi \left\{ -x_1+2x_2(1-x_1-x_3) \right\} \phi_K^A(x_2)
          \left( \phi_{\phi}^s(x_3) + \phi_{\phi}^t(x_3) \right)
      \right. \nonumber \\
      & & \;\;\;\;\;\;\;\; \left.
    - r_K r_\phi x_2 (1-x3)
          \left( \phi_K^P(x_2) + \phi_K^T(x_2) \right)
          \left( 3 \phi_{\phi}^s(x_3) + \phi_{\phi}^t(x_3) \right)
      \right. \nonumber \\
      & & \;\;\;\;\;\;\;\; \left.
    - 2 r_K r_\phi x_2
          \left( \phi_K^P(x_2) + \phi_K^T(x_2) \right)
          \phi_{\phi}^t(x_3)
      \right. \nonumber \\
      & & \;\;\;\;\;\;\;\; \left.
    + 2 r_K r_\phi x_1 x_2
          \left( 3\phi_K^P(x_2)\phi_{\phi}^s(x_3) 
                   + \phi_K^T(x_2)\phi_{\phi}^t(x_3) \right)
    \right. \nonumber \\
    & & \;\;\;\;\;\;\; \left.
    - 2 r_K r_\phi x_1
          \left( \phi_K^P(x_2) - \phi_K^T(x_2) \right) \phi_{\phi}^t(x_3)
\right]
\nonumber \\
& & \times 
E_{g}(t) h_n^{MP}(A,B,C,b_3,b_2,b_1)
\;,\\
%
%------------------------------------------------------------------
%
{\cal M}^{MP}_i
&=&
- 4 M_B^4 \frac{\sqrt{2N_c}}{2N_c} C_FN_c
\int_0^1 dx_1 dx_2 dx_3\int_0^{\infty} b_1db_1 b_2db_2 b_3db_3 
\phi_B(x_1,b_1)
\nonumber \\
& & \times
\left[ 
  \phi_K^A(x_2) \phi_{\phi}(x_3)
    +r_K \left( \phi_K^P(x_2) - \phi_K^T(x_2) \right)\phi_{\phi}(x_3)
\right.
\nonumber \\
& & \;\;\;\;\;\;\; \left.
    - r_\phi \phi_K^A(x_2) \left(\phi_{\phi}^s(x_3)
                                 +\phi_{\phi}^t(x_3)\right) 
    + 4 r_\phi r_K \phi_K^P(x_2) \phi_{\phi}^s(x_3)
\right]
\nonumber \\
& & \times 
E_{g}(t)  h_i^{MP}(B,C,b_1,b_2,b_3)
\;,\\
%
%------------------------------------------------------------------
%
{\cal M}^{MP}_j
&=& 
2 M_B^6 \frac{\sqrt{2N_c}}{2N_c}  C_FN_c
\int_0^1 dx_1 dx_2 dx_3\int_0^{\infty} b_1db_1 b_2db_2 b_3db_3 
\phi_B(x_1,b_1)
\nonumber \\
& & \times
\left[
-2
\left\{
x_1x_2 - (2+x_2)x_3 
\right\} \phi_K^A(x_2)\phi_{\phi}(x_3)
\right.\nonumber \\
& & \;\;\;\;\;\;\; \left.
+ r_K \left\{ 3(2+x_1(-2+x_2)+x_3-x_2x_3)\phi_K^P(x_2)
\right.\right.\nonumber \\
& & \;\;\;\;\;\;\;\;\;\;\;\; \left.\left.
      + (2+x_1(2+x_2)-3x_3-x_2(4+x_3))\phi_K^T(x_2) \right\}\phi_{\phi}(x_3)
\right.\nonumber \\
& & \;\;\;\;\;\;\; \left.
+  r_\phi \left\{ x_1^2-2x_1(-3+x_2+3x_3)+x_3(-2+2x_2+5x_3) \right\}
          \phi_K^A(x_2)\phi_{\phi}^s(x_3)
\right.\nonumber \\
& & \;\;\;\;\;\;\; \left.
-  r_\phi (-2+3x_1-2x_2-3x_3)(x_1-x_3)\phi_K^A(x_2)\phi_{\phi}^t(x_3)
\right.\nonumber \\
& & \;\;\;\;\;\;\; \left.
+ 6 r_\phi r_K \left\{3 (-x_1+x_3)\phi_K^P(x_2) 
                     - (x_1-x_1x_2+x_2x_3)\phi_K^T(x_2) \right\}
   \phi_{\phi}^s(x_3)
\right.\nonumber \\
& & \;\;\;\;\;\;\; \left.
-2 r_\phi r_K \left\{ -3(x_1-x_1x_2+x_2x_3)\phi_K^P(x_2) 
                     \right. \right.\nonumber \\
    & & \;\;\;\;\;\;\; \left.\left.
    + (x_1(3+2x_2)-(1+2x_2)x_3)\phi_K^T(x_2) \right\}
   \phi_{\phi}^t(x_3)
\right] 
\nonumber \\
& & \times 
E_{g}(t) h_j^{MP}(A,B,C,b_1,b_2,b_3)
\;.
\end{eqnarray}
The hard scale $t$ is defined in Eq.~(\ref{eq:hs}), and the hard function
$h_e^{MP}$ is defined in Eq.~(\ref{eq:he}). The other hard functions are
defined as follows: 
\begin{eqnarray}
\lefteqn{h_n^{MP}(A,B,C,b_1,b_2,b_3)}\hspace{0mm}\nonumber\\
&=&
 -
\int_0^\infty b_1 d b_1 b_2 d b_2 b_3 d b_3
  \frac{1}{(2\pi)^3}
  \int_0^{2\pi} d\theta'_1 d\theta'_2 d\theta'_3
\nonumber \\
&& \times
    K_0 \left( C \sqrt{ b_1^2+b_2^2
                        +2b_1b_2\cos(\theta'_1-\theta'_2)}\right)
    K_0 \left( B \sqrt{ b_2^2
                        +b_3^2+2b_2b_3\cos(\theta'_2-\theta'_3)}\right)   
\nonumber \\
&& \times
\int_0^\infty dX \frac{X \cos (X|D|) +iX\sin (XD) }{X^2+A^2}  
\;,
\\
\lefteqn{
h_i^{MP}(B,C,b_1,b_2,b_3)
 }\hspace{0mm}\nonumber\\
&=&
  K_0\left( B b_1 \right) 
  K_0\left( C b_3 \right) 
 \left\{
    \begin{array}{ll}
    1/(2\pi \Delta)
    &    \mbox{for}\ \  b_1+b_2 > b_3 >|b_1-b_2| \\
    0 &  \mbox{otherwise}
  \end{array}\;, 
\right.
\\
\lefteqn{h_j^{MP}(A,B,C,b_1,b_2,b_3)}\hspace{0mm}\nonumber\\
&=&
-
\int_0^{2\pi} d\theta'_1 d\theta'_2
\frac{
f\left(
b_1, b_2, \sqrt{b_1^2 + b_2^2 +
2b_1b_2\cos(\theta_1-\theta_2)} \right)
}
{(2\pi)^2
C^2}
\nonumber \\
&& \hspace{20mm} \times
  K_0( Ab_1)  
  K_0\left( B
         \sqrt{b_1^2 + b_2^2 + 2b_1b_2\cos(\theta_1-\theta_2)}\right)
\;.
\end{eqnarray}
Here,  
$\Delta = \sqrt{2b_1^2b_2^2+2b_2^2b_3^2
+2b_3^2b_1^2-b_1^4-b_2^4-b_3^4}/4$ and 
$ D \equiv b_1\cos\theta'_1 + b_2\cos\theta'_2 + b_3\cos\theta'_3$.
The quantities
$A$, $B$ and $C$, which are functions of $x_1$, $x_2$ and $x_3$, are
given in Table~\ref{tab:ABC2}. 
For simplicity, we ignored $|\boldsymbol{k}_{1T}-\boldsymbol{k}_{3T}|$ in the
calculation of $h_j^{MP}$. 

\begin{table}[hbt]
\caption{The definitions of $A$, $B$ and $C$}
\label{tab:ABC2}
\begin{center}
\begin{tabular}{c|ccc} 
\hline\hline
Diagram & $A$ & $B$ & $C$\\
\hline
(a) & $\sqrt{x_2}M_B$ & $\sqrt{x_1x_2}M_B$ & 
      $ i \sqrt{(1-x_2)x_3}M_B$ \\
(b) & $\sqrt{x_1}M_B$ & $\sqrt{x_1x_2}M_B$ & 
      $ \sqrt{x_1-x_3}M_B$ \\
(c) & $i \sqrt{x_2(1-x_1-x_3)}M_B$ & 
      $\sqrt{x_1x_2}M_B$ & 
      $i \sqrt{(1-x_2)x_3}M_B$ \\
(d) & $\sqrt{x_2(x_1-x_3)}M_B$ 
      & $\sqrt{x_1x_2}M_B$ &
      $\sqrt{x_1-x_3}M_B$ \\
(e) & $i \sqrt{x_3}M_B$ & $ i \sqrt{(1-x_2)x_3}M_B$ & 
      $\sqrt{x_1-x_3}M_B$ \\
(f) & $ i \sqrt{1-x_2}M_B$ & $ i \sqrt{(1-x_2)x_3}M_B$ & 
      $\sqrt{x_1x_2}M_B$ \\
(g) & $\sqrt{(x_1-x_3)(1-x_2)}M_B$ & 
      $ i \sqrt{(1-x_2)x_3}M_B$ & 
      $\sqrt{x_1-x_3}M_B$ \\
(h) & $\sqrt{1-x_2(1-x_1-x_3)}M_B$ & $ i \sqrt{(1-x_2)x_3}M_B$ & 
      $\sqrt{x_1x_2}M_B$ \\
(i) & ------  & $\sqrt{x_1x_2}M_B$ & 
      $ i \sqrt{(1-x_2)x_3}M_B$ \\
(j) & $\sqrt{x_1x_2}M_B$ & $ i \sqrt{(1-x_2)x_3}M_B$ & 
      $\sqrt{x_1-x_3}M_B$ \\
\hline
\end{tabular}
\end{center}
\end{table}
%

%-------------------------------------------------------------------%
%
%                  References
%
%-------------------------------------------------------------------%

\end{document}